\documentclass[twocolumn, superscriptaddress, showpacs, preprintnumbers, amsmath, amssymb]{revtex4}
\usepackage{graphicx}
\usepackage{bm}
\usepackage{mathrsfs}
\usepackage{dcolumn}

\begin{document}

\preprint{APS/123-QED}

\title{Controlling the nonlinear intracavity dynamics of large He-Ne laser gyroscopes}

\author{D. Cuccato}

\altaffiliation[Corresponding author: ]{davidecuccato@gmail.com}

\affiliation{Department of Information Engineering , University of Padova Via Gradenigo 6/B, I-35131 Padova, Italy.}

\affiliation{INFN National Laboratories of Legnaro, Viale dell'Universit\`a 2, I-35020 Legnaro, Padova, Italy.}

\author{A. Beghi}

\affiliation{Department of Information Engineering , University of Padova Via Gradenigo 6/B, I-35131 Padova, Italy.}

\affiliation{INFN Sezione di Pisa, Largo Bruno Pontecorvo 3, I-56127 Pisa, Italy.}

\author{J. Belfi}

\affiliation{Department of Physics, E. Fermi, University of Pisa Largo Bruno Pontecorvo 3, I-56127 Pisa, Italy.}

\affiliation{INFN Sezione di Pisa, Largo Bruno Pontecorvo 3, I-56127 Pisa, Italy.}

\author{N. Beverini}

\affiliation{Department of Physics, E. Fermi, University of Pisa Largo Bruno Pontecorvo 3, I-56127 Pisa, Italy.}

\affiliation{INFN Sezione di Pisa, Largo Bruno Pontecorvo 3, I-56127 Pisa,
Italy.}

\author{A. Ortolan}

\affiliation{INFN National Laboratories of Legnaro, Viale dell'Universit\`a
2, I-35020 Legnaro, Padova, Italy.}

\author{A. Di Virgilio}

\affiliation{INFN Sezione di Pisa, Largo Bruno Pontecorvo 3, I-56127 Pisa,
Italy.}

\date{\today}

\begin{abstract}
A model based on Lamb's theory of gas lasers is applied to a He-Ne ring laser gyroscope to estimate and remove the laser dynamics contribution from the rotation measurements. The intensities of the counter-propagating laser beams exiting one cavity mirror are continuously observed together with a monitor of the laser population inversion. These observables, once properly calibrated with a dedicated procedure, allow  us to estimate cold cavity and active medium parameters driving the main part of the nonlinearities of the system. 
The parameters identification and noise subtraction procedure has been verified by means of a Monte Carlo study of the system, and experimentally tested on the G-PISA ring laser oriented with  the normal to the ring plane almost parallel to the Earth rotation axis. 
In this configuration the Earth rotation-rate provides the maximum Sagnac effect while the 
contribution of the orientation error is reduced at minimum.  
After the subtraction of laser dynamics by a Kalman filter, the relative systematic errors of G-PISA reduce from 50 to $5$ parts in $10^{3}$ and can be attributed to the residual uncertainties on geometrical scale factor and orientation of the ring.
\end{abstract}

\pacs{05.45.-a, 42.55.Lt, 06.30.Gv}

\maketitle

\section{Introduction\label{sec:Introduction}}

Modern Ring Lasers gyroscopes (RL) based on the Sagnac effect are standard inertial sensors 
of rotation rates with many applications, ranging form inertial guidance \cite{application}, to angle 
metrology \cite{application1}, geodesy \cite{application2, ullirev}, and geophysics \cite{ullirev, application3}. Their application to General Relativity tests has also been recently considered \cite{application4} as in the GINGER proposal (Gyroscopes In General Relativity), an underground experiment on the detection of the dragging of the inertial frame (Lense-Thirring effect) due to the Earth rotation. A ring laser ``gyro'' consists of two counter propagating electromagnetic waves along a polygonal closed path that acts as a resonant cavity. The theory of Sagnac effect predicts that the resonance frequencies for the two counter-propagating modes inside a ring laser cavity, rotating with respect to an inertial frame, differ by 
\begin{equation}
\omega_{s}=\frac{8\pi A}{\lambda L}\,\omega_{r},\label{eq:stmodel1}
\end{equation}
where $\lambda$ is the wavelength of the laser beam, $\omega_{r}=\mathbf{n}\cdot\mathbf{\Omega}$ is the projection of the angular velocity of the cavity $\mathbf{\Omega}$ (relative to an inertial frame) on the normal vector $\mathbf{n}$ of the plane of the ring cavity, and $A$ and $L$ are its area and perimeter, respectively. Laser non-linearities, cavity non-reciprocities, and back-scattering of light on cavity mirrors, require the application of some corrections to the ideal case of Eq.(\ref{eq:stmodel1}) and we have to generalize it with  the effective formula \cite{Aronowitz}

\begin{equation}
\omega_{s}=\frac{8\pi A}{\lambda L}(1+\delta_{s})\,\omega_{r}+\Delta\omega+\Delta\omega_{0},\label{eq:stmodel1-1}
\end{equation}
where $\Delta\omega_{0}$ gives the null shift error from the zero value of ${ \displaystyle \mathbf{\Omega}}$, $\Delta\omega$ accounts for the non-linear contributions from laser dynamics and two beams coupling, and $\delta_{s}$ represents the scale factor modifications produced by fluctuations of temperature, pressure and gain of the plasma, as well as by fluctuations of losses, and laser frequency. An exhaustive treatment of gas lasers phenomena is given by E. Lamb in ref. \cite{Lamb}. He described  the coupling between the laser beams and the atomic polarization at the third-order expansion in the interaction field, introducing a system of differential equations ruled by a set of parameters known in the following literature as ``Lamb parameters". 

The accurate knowledge  of scale factor and null shift is particularly required in tests of fundamental physics with RLs, e.g. the local Lorentz invariance, Axion detection, frame dragging, etc. \cite{StedmanReport}. The motivation for our work is to foresee a calibration procedure for large ring lasers that exploits the non-linearity of laser active medium without the need for rotating the apparatus through a calibrated angle. The possibility of performing a  calibration without knowing the input signal is a typical feature of non-linear systems with non-homogeneous input-output relations. With the separation of the identification of cavity losses  and active medium parameters we have, on the one hand, that we can increase the time stability of a RL with Kalman filtering by subtracting the backscattering. On the other hand, with the characterization of laser active medium (single pass gain and plasma dispersion function), we can achieve a good accuracy of the rotation estimate. In this respect, we already developed  \cite{Noi} an algorithm for the identification of the Lamb parameters for excess gain minus losses $\alpha_{1,2},$ backscattering amplitude $r_{1,2}$ and phases $\varepsilon_{1,2}$, that are associated with cavity dissipation. After providing a raw estimate of the laser medium gain, we run an extended Kalman filter that is able to remove a fraction of the backscattering induced drift from the Earth rotation rate measurements. Thus we improved the long term stability of the ring G-PISA \cite{Noi,G-PISA,G-PISA1} and demonstrated the effectiveness of non-linear Kalman filtering in removing backscattering effects. In this paper we extend our previous analysis by presenting a novel identification and calibration procedure for estimating the full set of Lamb parameters, i.e. $\alpha_{1,2},$ $r_{1,2},$ $\varepsilon_{1,2},$ self- and cross- saturation coefficients $\beta_{1,2}$ and $\theta_{12,21}$, and scale factor and null shift errors $\sigma_{1,2}$ and $\tau_{12,21}$, which depend in their turn on cavity losses $\mu_{1,2}$, plasma polarizability and single pass gain $G$ of the laser medium. 

The paper is organized as follows. In Sect. \ref{sec:Ring-Laser-Dynamics} we report the basic equations of the ring laser dynamics. With the help of a Monte Carlo simulation , we show in Sect. \ref{sec:Calibration-Procedures} the required bound on relative
errors of Lamb parameters as a function of the accuracy goal in the estimation of rotation rate. The experimental apparatus of G-PISA and the calibration setup are presented in Sec. \ref{sec:Experimental-Apparatus}. In Sec. \ref{sec:Results-and-Discussion} we apply our calibration procedure to G-PISA data and show the results. In Sec. \ref{sec:Conclusions} conclusions are drawn, limitations of calibration and identification procedures, including calibration goals for application of RL to fundamental physics, are discussed.

\section{Ring laser dynamics and cold cavity parameters estimation} \label{sec:Ring-Laser-Dynamics}

To study the RL dynamics we refer to the self-consistent equations derived by Aronowitz \cite{Aronowitz} starting from Lamb's third order expansion of the polarization vector in powers of the electric field amplitude. The standard equations of a RL with backscattering can be conveniently written using the complex representation of the optical fields, expressed in Lamb units.

\begin{equation}
\begin{array}{cc}
\dot{E_{1}}(t)= & \left[\mathcal{A}_{1}-\mathcal{B}_{1}|E_{1}(t)|^{2}-\mathcal{C}_{21}|E_{2}(t)|^{2}\right]\, E_{1}(t)+\mathcal{R}_{2}E_{2}\\
\dot{E_{2}}(t)= & \left[\mathcal{A}_{2}-\mathcal{B}_{2}|E_{2}(t)|^{2}-\mathcal{C}_{12}|E_{1}(t)|^{2}\right]\, E_{2}(t)+\mathcal{R}_{1}E_{1}
\end{array}\ ,\label{eq:2.1}
\end{equation}

where $E_{1,2}(t)$ and $\Omega_{1,2}$ are the complex amplitudes and frequencies of the clock-wise (identified by subscript 1) and counter-clock-wise (identified by subscript 2) waves propagating
in the cavity, and the complex coefficients $\mathcal{A}_{1,2},\,\mathcal{B}_{1,2},\,\mathcal{C}_{21,12},$ and $\mathcal{R}_{1,2}$ are related to the Lamb parameters $\alpha_{1,2},$ $\sigma_{1,2},$ $\beta_{1,2},$ $\theta_{12,21},$ $\tau_{12,21},$ $r_{1,2}$ and $\varepsilon_{1,2}$ by 

\begin{equation}
\begin{cases}
\mathcal{A}_{1,2}=\frac{c}{L}\frac{\alpha_{1,2}}{2}+i\left(\Omega_{1,2}-\omega_{ref}+\frac{c}{L}\sigma_{1,2}\right) & \ \\
\mathcal{B}_{1,2}=\frac{c}{L}\beta_{1,2}\\
\mathcal{C}_{21,12}=\frac{c}{L}\left(\theta_{21,12}-i\tau_{21,12}\right) & \ ,\\
\mathcal{R}_{1,2}=\frac{c}{L}r_{1,2}e^{i\epsilon_{1,2}}
\end{cases}\label{eq:2.2}
\end{equation}

where $\omega_{ref}$ is the rotation rate of the detector reference frame with respect to the local inertial frame.

By introducing the 2-dimensional complex valued column vector $\mathbf{E}(t) = \left( E_{1} , E_{2} \right) $, Eqs.(\ref{eq:2.1}) can be written as 

\begin{equation}
\mathbf{\dot{E}}=\left[A-\mathcal{\mathscr{D}}(\mathbf{E}) B \mathscr{D}(\mathbf{E^{\textrm{*}}\textrm{)}}\right]\mathbf{E}\:,\label{eq:2.3}
\end{equation}

where we have defined the complex-valued matrices

\begin{equation}
A \equiv\left(\begin{array}{cc}
\mathcal{A}_{1} & \mathcal{R}_{2}\\
\mathcal{R}_{1} & \mathcal{A}_{2}
\end{array}\right) \ 
B \equiv\left(\begin{array}{cc}
\mathcal{B}_{1} & \mathcal{C}_{21}\\
\mathcal{C}_{12} & \mathcal{B}_{2}
\end{array}\right),\label{eq:2.4}
\end{equation}

and

\begin{equation}
\mathscr{D}(\mathbf{E})\equiv\left(\begin{array}{cc}
E_{1} & 0\\
0 & E_{2}
\end{array}\right).\label{eq:2.5}
\end{equation}

In standard ring laser operation the rotational frequency is much smaller than the cavity modes spacing and we have $\mathcal{B}_{1}\simeq\mathcal{B}_{2}=\mathcal{B},$ $ \mathcal{C}_{21} \simeq \mathcal{C}_{12} = \mathcal{C}$. We will use this approximations throughout the paper when we neglect the subscript $1,2$ in the corresponding Lamb parameters. For instance, the relative difference is of the order of  $10^{-6}$ for large ring lasers biased by Earth rotation \cite{MM1}. 

It is worth noticing that analytical solutions of Eqs.(\ref{eq:2.3}) have been found in the case of $|{E_{1}}|\varpropto |{E_{2}}|$ (Adler Solution) \cite{Stedman}, $\mathcal{C}_{21}=\mathcal{C}_{12}$
and $\mathcal{R}_{1}=\mathcal{R}_{2}^{*}$ \cite{Pesquera}, or $Re[\mathcal{C}_{21}] = Re[\mathcal{C}_{12}] = \mathcal{B}_{1} = \mathcal{B}_{2},$ \cite{Chiba}. However, general analytic solutions of the RL equations system with non reciprocal parameters are not known, and therefore perturbative solutions have to be used \cite{Noi}.

By introducing the new coordinates $ \mathbf{I}= $ $ \left( E_{1}^{*}E_{1}, E_{2}^{*}E_{2} \right)^{T},$ and $X=\log\left(E_{1}/E_{2}\right),$ Eqs.(\ref{eq:2.3}) can be written as

\begin{equation}
{\displaystyle \begin{cases}
\frac{d}{dt}\log(\mathbf{I})  =2 \text{Re}\left[\left(\begin{array}{c}
\mathcal{A}_{1}\\
\mathcal{A}_{2}
\end{array}\right)- B \mathbf{I} + \left(\begin{array}{c}
\mathcal{R}_{2}e^{-X}\\
\mathcal{R}_{1}e^{X}
\end{array}\right)\right]\\
\frac{d}{dt}X  =\mathcal{A}_{1}-\mathcal{A}_{2}-(\mathcal{B}-\mathcal{C})(I_{1}-I_{2})+\mathcal{R}_{2}e^{-X}+\mathcal{R}_{1}e^{X}
\end{cases}}\label{eq:2.6}
\end{equation}

which reduce to Eqs. (6) of ref. \cite{Noi} after one identifies $I_{1} = E_{1}^{*} E_{1}$, $I_{2} = E_{2}^{*} E_{2}$, and $\psi=Im\left[X\right],$ and further assumes that $\mathcal{C}=0$. In the reciprocal case (i.e. $\mathcal{A}_{1}=\mathcal{A}_{2},$ $\mathcal{B}_{1}=\mathcal{B}_{2},$ $\mathcal{C}_{21} = \mathcal{C}_{12},$ and $\mathcal{R}_{1} = \mathcal{R}_{2}$), Eq. \ref{eq:2.3} is clearly invariant under the transformation $\left(E_{1},\, E_{2}\,,\Omega_{1},\Omega_{2}\right)$$\longrightarrow $$\left(E_{2},\, E_{1}\,,\Omega_{2},\Omega_{1}\right),$ and so the corresponding phase portrait is topologically equivalent to a torus \cite{reciprocal}. The system exhibits two asymptotic time behaviors depending on whether or not the orbits in the phase space can be continuously shrunk to a point. For behaviors of the first kind, Eqs.(\ref{eq:2.3}) exhibit a fixed point, the beat frequency is equal to zero and the light intensities are constant (laser switched off or locked-in). Conversely, the behaviors of the second kind are limit cycles regimes, characterized by nonzero beat frequency of the the counter-propagating waves (single mode laser operation). When the reciprocal conditions do not hold, the qualitative characteristics
of asymptotic time behavior are left unchanged, as a result of the invariance of phase portrait topology under continuous variation of RL parameters.

The RL equations depend on two sets of parameters with distinct physical origin: \emph{i)} cold cavity parameters associated to losses, and \emph{ii)} active medium parameters associated to atomic polarizability.  According to this classification, the ring laser matrices can be naturally written as

\begin{equation}
A \equiv\frac{c}{L} P^{(0)}- M \quad,\label{eq:3.1}
\end{equation}

\begin{equation}
 B \equiv\frac{c}{L} P^{(2)}\label{eq:3.2}
\end{equation}

where $ P^{(0)}$ and $ P^{(2)}$ are the $0$-th and $2$-nd order contributions to the gas mixture polarizability \cite{Lamb}

\begin{equation}
\begin{array}{c}
P^{(0)}={\displaystyle \frac{G}{2}}\left(\begin{array}{cc}
z^{(0)}(\xi_{1}) & 0\\
0 & z^{(0)}(\xi_{2})
\end{array}\right), \\
\\
\quad P^{(2)}=G\left(\begin{array}{cc}
z_{s}^{(2)}(\xi_{1}) & z_{c}^{(2)}(\xi_{1,2})\\
z_{c}^{(2)}(\xi_{1,2}) & z_{s}^{(2)}(\xi_{2})
\end{array}\right)\quad,\end{array}\label{eq:3.3}
\end{equation}

and $ M $ is the linear coupling matrix associated to cavity losses

\begin{equation}
 M= \left(\begin{array}{cc}
\frac{c}{L}\frac{\mu_{1}}{2}+i\omega_{1} & -\frac{c}{L}r_{2}e^{i\epsilon_{2}}\\
-\frac{c}{L}r_{1}e^{i\epsilon_{1}} & \frac{c}{L}\frac{\mu_{2}}{2}+i\omega_{2}
\end{array}\right)\quad;\label{eq:3.boh}
\end{equation}

here $ \omega_{1,2} = \Omega_{1,2}-\omega_{r} +\frac{c}{L}\sigma_{1,2},$ $G$ is the laser single pass gain, $\mu_{1,2}$ are the cavity losses, and $z^{(0)}(\xi_{1,2}) = (\alpha_{1,2} + \mu_{1,2})/G$, $z_{s}^{(2)}(\xi_{1,2}) = \beta_{1,2}/G$, $z_{c}^{(2)}  (\xi_{1,2})=\theta_{12,21}/G$ are functions which depend on the normalized detuning  $\xi_{1,2}$ of the optical frequencies to the cavity center frequency, and will be explicitly calculated in the next Section. Here we only recall that, in the case of large size RLs sensing the Earth rotation, we have $\xi_{1}-\xi_{2} < 10^{-6}$, and so we can approximate the frequency detuning of each beam with its average  $\xi=(\xi_{1}+\xi_{2})/2$. Consequently, cross and self saturation parameters can be assumed equal for the two beams, i.e. $\theta_{12} = \theta_{21} \equiv \theta$, $\tau_{12} = \tau_{21} \equiv \tau$ and $\beta_{1} = \beta_{2} \equiv \beta$.  

As mirror losses of an optical cavity are unpredictable, they must be identified in the RL outputs and tracked in time. To this aim, we assume for the moment that the first and second order polarizations $P^{(0)}$ and $P^{(2)}$ are given, and that the light intensities have been already calibrated in Lamb units (see Sect. 3). The identification procedure of  cold cavity parameters is based on the existence of a limit cycle that makes it possible to estimate the steady state dynamical losses of a RL system,  as asymptotic solutions are periodic with period $T=2\pi/\omega$.  In fact, from Eq.(\ref{eq:2.6}) we can construct the functional

\begin{equation}
J(\mathbf{I},X,M)\equiv \label{eq:2.7}
\end{equation}
\begin{equation*}
\left\Vert \frac{d}{dt}\log(\mathbf{I})-2\text{Re}\left[\left(\begin{array}{c}\mathcal{A}_{1}\\
\mathcal{A}_{2}
\end{array}\right)- B \mathbf{I}+\left(\begin{array}{c}
\mathcal{R}_{2}e^{-X}\\
\mathcal{R}_{1}e^{X}
\end{array} \!\!\!\right)\right]\right\Vert^{2}_{L^2(T)}\!,
\end{equation*}

where $\left\Vert \cdot\right\Vert_{L^2(T)} \equiv\sqrt{\int_{0}^{T}(\cdot)^{2}dt}$
is the norm in the Hilbert space of $L^{2}$ T-periodic signals, and search for its minimum value to derive statistics of parameters estimation. From the perturbative solutions in Eqs.(9) of ref. \cite{Noi} approximated up to the first harmonic terms, we can write the steady state intensities and complex phase difference of the electrical fields  as

\begin{equation}
\begin{cases}
I_{1}(t) & =I_{1}+i_{1}\sin\left(\omega t+\phi_{1}\right)\\
I_{2}(t) & =I_{2}+i_{2}\sin\left(\omega t+\phi_{2}\right)\\
X(t) & =\frac{1}{2}\log\left[\frac{I_{1}(t)}{I_{2}(t)}\right]+i\omega t
\end{cases}\quad , \label{eq:2.8}
\end{equation}

where $I_{1,2}$, $i_{1,2}$, and $\phi_{1,2}$ are the intensity offsets, monobeam modulation amplitudes and phases, which can be readily measured from RL outputs \cite{Noi}; it also follows that the backscattering phases can be estimated by $\widehat{\epsilon_{1}}=\phi_{1}$ and $\widehat{\epsilon_{2}} = \phi_{2}$ modulo a common initial phase. Thus, after substituting Eqs.(\ref{eq:2.8}) into Eq.(\ref{eq:2.7}), we can estimate the cavity loss parameters $\mu_{1,2}$ and $r_{1,2}$ as

\[
\left(\hat{\mu}_{1,2},\hat{r}_{1,2}\right)=\text{argmin}_{\mu_{1,2},r_{1,2}}\left\{ J(I,X, M)\right\} .
\]

Assuming all the other parameters known, the identified cold cavity parameters read

\begin{equation}
\begin{cases}
\hat{\mu}_{1,2} & =\alpha_{0}-\beta\left(I_{1,2}+\frac{i_{1,2}^{2}}{I_{1,2}}\right)-\frac{i_{1}i_{2}I_{2,1}(L\omega/c) \text{cos}\widehat{\epsilon}}{4I_{1,2}^{2}} - \\
& -\theta\left(\frac{i_{2,1}^{2}+4I_{2,1}^{2}}{4I_{1,2}}-\frac{i_{1,2}^{2} I_{2,1}^{2}}{2I_{1,2}^{3}}+\frac{i_{1}i_{2}I_{2,1}\cos\widehat{\epsilon}}{I_{1,2}^{2}}+\frac{i_{2,1}^{2}\cos2\widehat{\epsilon}}{4I_{1,2}}\right) \\
\hat{r}_{1,2} & =\frac{i_{2,1}(L\omega/c)}{2\sqrt{I_{1}I_{2}}}\mp i_{1,2}\sqrt{\frac{I_{1,2}}{I_{2,1}}}\theta\sin\widehat{\epsilon}
\end{cases}.\label{eq:2.9}
\end{equation}

where $\widehat{\epsilon}=\widehat{\epsilon}_{1}-\widehat{\epsilon}_{2},$ and $\alpha_{0}$ is the excess gain for zero losses. Eqs.(\ref{eq:2.9}) reduce to Eqs.(13-16) of ref.~\cite{Noi} if the cross-saturation parameter $\theta=0$ and $\alpha_{0}=G$ (Doppler limit approximation). As we have already shown, we can improve the long term stability of the estimated Sagnac frequency by using Extended Kalman filters (EKF) \cite{Noi}.

However, to provide and improve also the accuracy of the Sagnac frequency, unbiased estimates of the active medium parameters $\beta$, $\theta$, and $\tau$, are required. Unfortunately, the minimization of the functional in Eq.(\ref{eq:2.7}) is ill-posed for the full set of Lamb parameters, and so different approaches for their measurement must be investigated. 

\section{Active medium analysis \label{sec:Calibration-Procedures}}
Theoretical and experimental results of laser plasma physics allow us to estimate the active medium parameters by changing the working point of the system, inducing the ``multimode transition", and  monitoring the plasma fluorescence \cite{MM,MM1}.  A RL can be so calibrated as an inertial rotation sensor. However, to get an order of magnitude of the required accuracy on the active medium parameters given a target precision on the accuracy of the Sagnac frequency, we start the analysis of the calibration problem numerically by Monte Carlo techniques. Then we focus on the description of atomic polarization of the active laser medium and  the corresponding experimental measurements.

\subsection{Monte Carlo simulations}

Firstly, we study the bias induced in the estimate of the Sagnac frequency  by the approximations made in Eq.(\ref{eq:2.9}). We use the typical parameters of G-PISA \cite{Noi} as given in Tab. \ref{tab:refGP}, and a step size of $0.2$ msec to integrate the Eqs.(\ref{eq:2.6}). The parameters $\alpha_{1,2}$, $r_{1,2}$ and $\varepsilon$ are simulated as independent random walks with starting value as in Tab.\ref{tab:refGP}, correlation time of half of the simulation length $ 300$ s, and relative step size of $10^{-2}$. The initial value of the backscattering phase $\varepsilon$ is assumed uniformly distributed in $[0 ,\pi/2]$. We find that the relative accuracy of the Sagnac frequency estimation is few parts in $10^5$, with relative systematic errors of $10^{-4}$ and $10^{-2}$ on the identified parameters $\hat{\mu}_{1,2}$ and $\hat{r}_{1,2}$, respectively. The phase errors are $\sim 10^{-3}$ rad. 

\begin{table}[h]
\begin{tabular}{|c|c|c|}
\hline 
$c/L$ & $5.5\times10^{7}$Hz \\ 
\hline
$\alpha_{1,2}$ & $\sim10^{-6}$ \\
\hline
$\beta$ & $5\times10^{-5}$ \\
\hline
$\theta$ & $6.5\times10^{-6}$ \\
\hline
$r_{1,2}$ & $\sim2\times10^{-7}$ \\
\hline
$\tau$ &  $180$ rad/s \\
\hline 
\end{tabular}
\caption{\label{tab:refGP} Reference values of the Lamb parameters used in the simulation of G-PISA dynamics. The contribution of  $r_{1,2}$ and $\tau$ to the Sagnac frequency is of the order of $1.5$ Hz and $10^{-2}$ Hz, respectively.}
\end{table}

Moreover, the Monte Carlo simulations show that the precision and accuracy of the identification procedure increases with the dimension of the ring, as  the values of Lamb parameters decrease linearly with the free spectral range; e.g. for a square ring with a side of $\simeq 1$, $5$ and $10$ m  we found that the relative frequency accuracy is few parts in $10^5$, $10^8$, and $10^9$, respectively. Other Monte Carlo simulations have been run by appropriately biasing the intensity time series, to mimic a systematic error in their calibration in Lamb units or acquisition process. It results that the corresponding relative error in the Sagnac frequency estimation mainly depends on the values of $r_{1,2}$, $\tau$ and $\varepsilon$. In particular, for the typical G-PISA parameters in table \ref{tab:refGP}, the relative frequency error scales linearly with the intensity error, and it turns out to be $\simeq 1$ part in $10^4$ times the intensity relative error. Finally, we studied the effects of systematic errors of the laser active medium parameters. The end result of these Monte Carlo runs is that the accuracy of the Sagnac frequency estimation is dominated by systematic errors of the single pass gain. For instance, by biasing the relevant active medium parameters $\beta$, $\theta$ and $\tau$ of a relative error of  $10^{-1}$, $10^{-2}$, and $10^{-3}$, the corresponding relative error in the Sagnac frequency turns out to be  $10^{-4}$, $10^{-5}$, and $10^{-6}$, respectively. 

\subsection{2$^{nd}$-Order calculation of Plasma Polarization}

 The complex valued functions $z^{(0)}(\xi_{1,2})$,  $z_{s}^{(2)}(\xi_{1,2})$, and
$z_{c}^{(2)}(\xi_{1,2})$ which allow us to estimate self- and cross- saturation parameters are rather common in plasma physics, and were calculated for the first time by Aronowitz with two counter-propagating laser beams \cite{Aronowitz,Aronowitz1}. His model of  plasma requires that the ratio $\eta$ between homogeneous $\gamma_{ab}$ and inhomogeneous $\Gamma$ broadening line width is $\eta\equiv\gamma_{ab}/\Gamma \ll1$. For instance, in the experiment of ref. \cite{Aronowitz1}, the typical value of $\eta$ is $\simeq 10^{-2}$. For gas mixtures with pressure $4 \div 8$ mbar and temperature $300 \div 500$ K, as in G-PISA, we have instead $0.2 \le \eta \le  0.5$ \cite{Noi}, and so we must perform a more general calculation of the atomic polarization. However, we will follow the approach of Aronowitz in ref.  \cite{Aronowitz} for what concerns the series expansion in powers of the electric fields describing the interaction between radiation and atoms. Aronowitz showed that the complex polarization of the active medium in a gas He-Ne laser, expanded to the third order in the field amplitude, can be written in the following integral form 

\begin{equation}
\mathcal{ P}^{(3)}(E_{1,2})=-\frac{2i|\mu_{ab}|^{2} E_{1,2}}{\gamma_{ab}}\int_{0}^{\infty}\chi^s_{1,2}(v)\,\rho^{(2)}(v,E_{1,2})\, dv\ ,\label{eq:A1}
\end{equation}

where $|\mu_{ab}|$ is the electric dipole moment between states $a$ and $b,$ $\gamma_{ab}$ is the homogeneous line width, $v$ is the velocity of atoms, $\chi^s_{1,2}(v)=1/(i\eta+\xi_{1,2}\pm v/u)$ is the complex susceptibility, $u$ is the atomic mass constant, and 

\begin{widetext}
\begin{equation}
\rho^{(2)}(v,E_{1,2})=\frac{N\, e^{-\frac{v^{2}}{\Gamma^{2}}}}{2\gamma_{a}\gamma_{b}\hbar\Gamma}\left(1-I_{1}\frac{1}{1+(\xi_{1}^{'}+v/u)^{2}}-\frac{\gamma_{a}+\gamma_{b}}{\gamma_{ab}}I_{2}\frac{1}{1+(\xi_{2}^{'}-v/u)^{2}}\right)\label{eq:A2}
\end{equation}
\end{widetext}

is the second order population inversion. Here $\gamma_{a}$ and $\gamma_{b}$ are the decay rates of the upper and lower energy level, $N$ is the average excitation inversion density, $I_{1,2}$ are the normalized light intensities of $E_{1,2}$ expressed in Lamb units, and $\xi_{1,2}^{'} = (\omega - \omega_{1,2})/\gamma_{ab}$ is the frequency detuning to the natural line width ratio, $\omega_{1,2}$ and $\omega$ are the optical and cavity frequencies, respectively. It is worth mentioning that the polarization depends on both amplitudes and frequencies of the electrical fields. After a decomposition in simple fractions of the rational part of Eq.(\ref{eq:A2}) , the integral in Eq.(\ref{eq:A1}) can be conveniently evaluated by means of the plasma dispersion function 

\begin{equation*}
\text{\ensuremath{\mathcal{Z}}}\left(y\right)\equiv\mathcal{Z}_{R}(y)+i\mathcal{Z}_{I}(y)= 
\end{equation*}
\begin{equation}
\frac{1}{\sqrt{\pi}}\int_{\mathbb{R}}\frac{e^{-x^{2}}}{x-i\eta-y}\, dx= \label{eq:A3}
\end{equation}
\begin{equation*}
i\sqrt{\pi}e^{-(\eta-iy)^{2}}\mathrm{erfc}\left(\eta-iy\right)\quad,
\end{equation*}

where $\mathrm{erfc}(\cdot)$ is the complementary error function, and $\mathcal{Z}_{R}$ and $\mathcal{Z}_{I}$ are the real and imaginary parts of the plasma dispersion function, and  are proportional to the spectral gain and dispersion profiles of the unsaturated active medium, respectively. 

The expression for the atomic polarization up to the 3-rd order approximation reads 

\begin{equation} \label{eq:A4}
\mathcal{ P}^{(3)}(E_{1,2}) \equiv  \chi(E_{1,2}) E_{1,2} =
\end{equation}
\begin{equation*}
\frac{\sqrt{\pi}A\mathcal{Z}_{I}\left(0\right)}{\gamma_{ab}\gamma_{a}\gamma_{b}}\left(z^{(0)}(\xi_{1,2})-z_{s}^{(2)}(\xi_{1,2})I_{1,2}-z_{c}^{(2)}(\xi)I_{2,1}\right)E_{1,2}\quad,
\end{equation*}

where $\chi(E_{1,2})$ is the cavity atomic polarizability, $A=N|\mu_{ab}|^{2}/(\hbar\Gamma),$ 

\begin{equation}
\begin{cases}
 z^{(0)}(\xi_{1,2}) =  \displaystyle{\frac{\mathcal{Z}\left(\xi_{1,2}\right)}{\mathcal{Z}_{I}\left(0\right)}}\\
 z_{s}^{(2)}(\xi_{1,2}) =  \displaystyle{\frac{\mathcal{Z}_{I}(\xi_{1,2})(1-2\eta(\eta+i\xi_{1,2}))}{\mathcal{Z}_{I}\left(0\right)}+}\\

\qquad \qquad \ + \displaystyle{\frac{2\eta-2\eta\mathcal{Z}_{R}(\xi_{1,2})\left(i\eta-\xi_{1,2}\right)}{\mathcal{Z}_{I}\left(0\right)}}  \\ \label{eq:A5}

z_{c}^{(2)}(\xi) =  \displaystyle{\frac{\gamma_{a}+\gamma_{b}}{\gamma_{ab}}\frac{\eta}{\xi}\frac{\overline{\mathcal{Z}}_{I}(\xi)\xi-\overline{\mathcal{Z}}_{R}(\xi)\eta-\overline{\mathcal{Z}}_{i}(\xi)(\xi-i\eta)}{\mathcal{Z}_{I}\left(0\right)(\eta+i\xi)}}
\end{cases},
\end{equation}

and $\xi_{1,2}=\pm(\omega-\omega_{1,2})/\Gamma$ is the frequency detuning to the Doppler width ratio, $\mathcal{\overline{Z}}_{I}=\left[\mathcal{Z}_{I}\left(\xi_{1}\right)+\mathcal{Z}_{I}\left(\xi_{2}\right)\right]/2,$
$\overline{\mathcal{Z}}_{R}=\left[\mathcal{Z}_{R}\left(\xi_{1}\right)+\mathcal{Z}_{R}\left(\xi_{2}\right)\right]/2,$
$\overline{\mathcal{Z}}_{i}=\left[\mathcal{Z}_{I}\left(\xi_{1}\right)-\mathcal{Z}_{I}\left(\xi_{2}\right)\right]/2$.
Using Eq.(\ref{eq:A4}), the expression of r.h.s. of Eq.(\ref{eq:3.3}), (\ref{eq:3.boh}) can be identified with the coefficients of $z^{(0)}(\xi_{1,2}),\, z_{s}^{(2)}(\xi_{1,2})$ and $z_{s}^{(2)}(\xi_{1,2})$ of Eq.(\ref{eq:A5}), except for the constant laser single pass gain $G$.

\begin{figure}
\includegraphics[width=8cm]{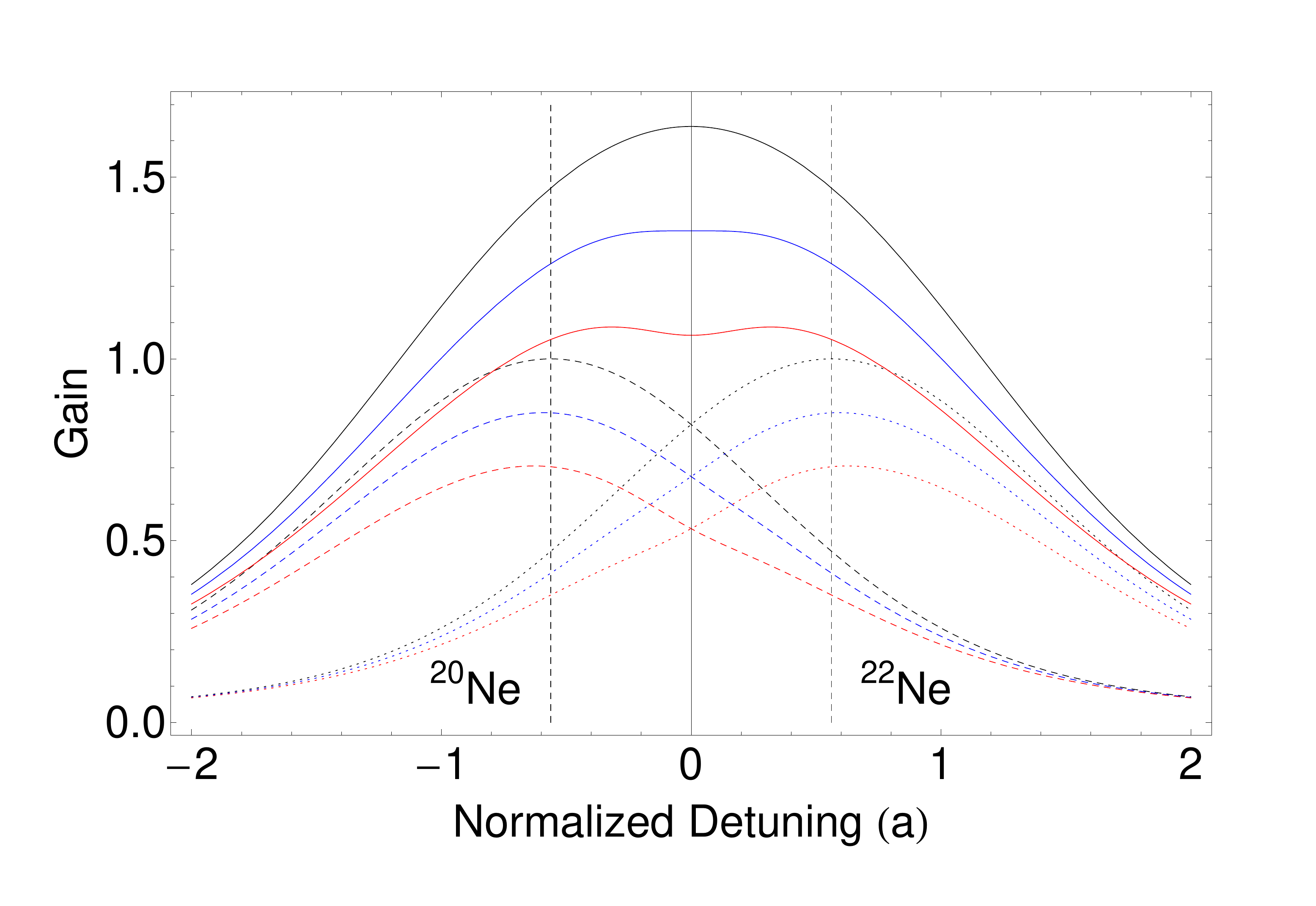}
\includegraphics[width=8cm]{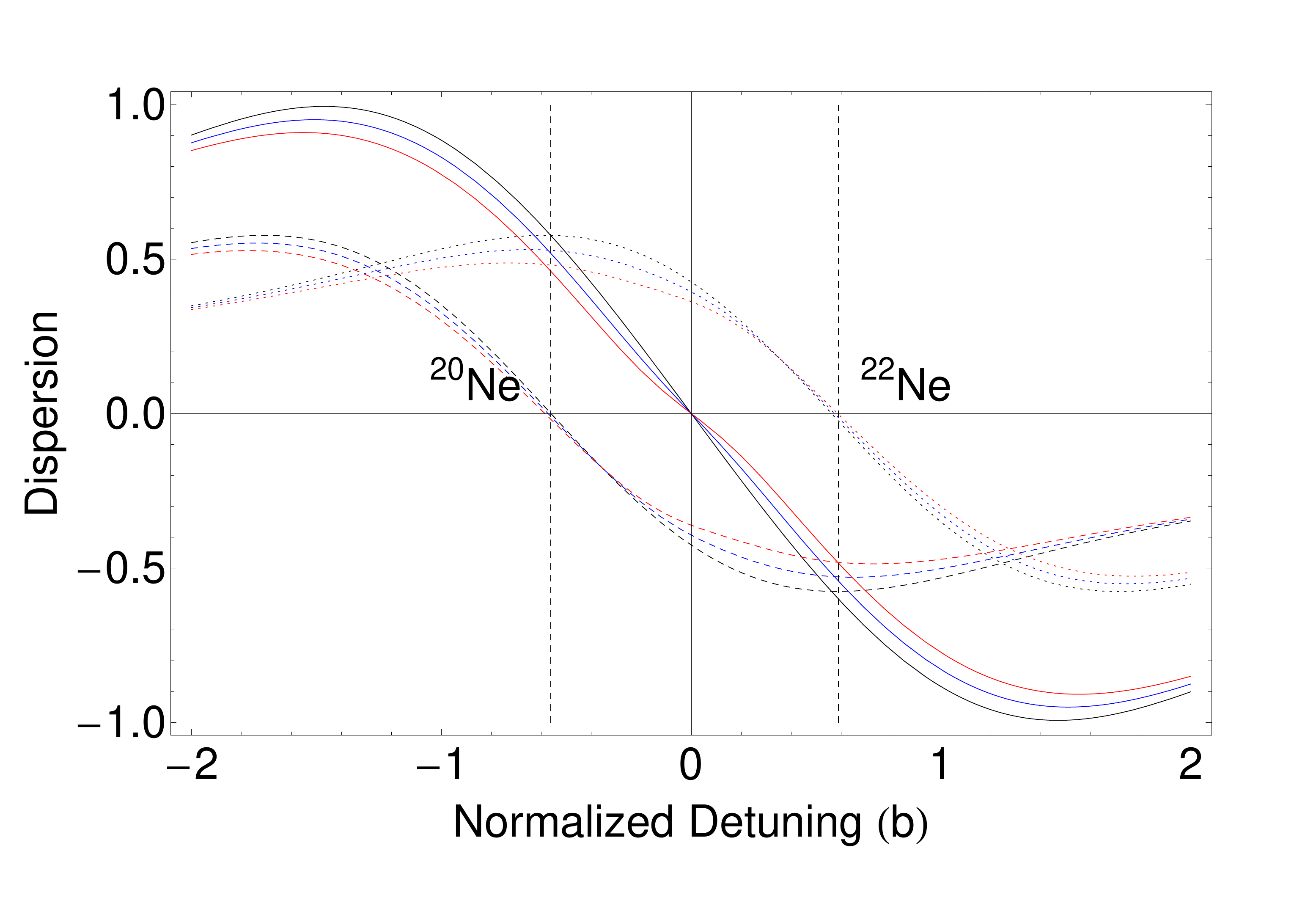}
\caption{\label{fig:A1} Plot of the computed gain (a) and dispersion (b) profiles of the plasma polarizability in the ring laser cavity (continuous line), and of its contributions from $^{20}$Ne isotope (dashed line), and  $^{22}$Ne isotope (dotted line), assuming $I_{1}=I_{2}=I$. The black lines represent the unsaturated profile ( $I \rightarrow 0$), the blue and red lines represent profiles saturated by signals of intensity $I=0.1$, and $I=0.2$, respectively. The vertical dashed lines indicate the centers of the broadening profiles of the $Ne$ isotopes.} 
\end{figure}

In our calculations we must take into account that usually a RL cavity is filled with a gas mixture of two Neon isotopes. Thus the matrix elements of Eq.(\ref{eq:3.3}) must be substituted by

\begin{equation}
\begin{cases}
z^{(0)}(\xi_{1,2})= & k^{'} z^{(0)} (\xi^{'}_{1,2})+k^{''} z^{(0)} (\xi^{''}_{1,2})\\
z_{s}^{(2)}(\xi_{1,2})= & k^{'}\, z_{s}^{(2)}(\xi_{1,2}^{'})+k^{''}\, z_{s}^{(2)}(\xi^{''}{}_{1,2})\\
z_{c}^{(2)}(\xi)= & k^{'}\, z_{c}^{(2)}(\xi^{'})+k^{''}\, z_{c}^{(2)}(\xi^{''})
\end{cases}
\quad,\label{eq:A6}
\end{equation}

where the symbols $\ ^{'}$ and $\ ^{''}$ refer to the $^{20}$Ne and $^{22}$Ne isotopes,  $k^{'}$ and $k^{''}$ are the fractional amount of each isotope, and $\xi_{1,2}^{'},\,\xi_{1,2}^{''}$ are the detuning to the center frequency of each isotope. Practically, the values of $\eta$, $\xi$ and $k$ are rescaled by the square root of the ratio of the atomic mass of the two isotopes. As an example, we show in Fig.\ref{fig:A1} the polarization of a 50-50 $^{20}$Ne-$^{22}$Ne gas mixture as the sum of contributions arising from each Ne isotope, according to Eq.(\ref{eq:A6}). 

\subsection{Round Trip Losses}

The spectroscopic technique known as ``Ring Down Time measurement'' (RDT) allows us to estimate mirror losses from the impulse response of a linear system. In fact, from Eq.(\ref{eq:2.3}) with $G=0,$ we have

\begin{equation}
\mathbf{\dot{E}}= M \mathbf{E}\ ,\label{eq:3.4}
\end{equation}

and so the system shows two exponential decays with a rate proportional to the total cavity losses.  If we switch off the laser excitation at the time $t=0,$ the initial conditions are $\mathbf{E}(0) = \left[ \sqrt{I_{1}(0)}e^{i\phi_{1}} , \sqrt{I_{2}(0)}e^{i\phi_{2}} \right],$ where $I_{1,2}(0)$ are the initial intensities, and $\phi_{1,2}$ are the initial phases. 

The solutions of Eq.(\ref{eq:3.4}) for the light intensities, expanded in a Taylor series of $\omega\gg\frac{c}{L}r_{1,2},\,\frac{c}{L}\mu_{1},\frac{c}{L}\mu_{2}$ to the first order, reads

\begin{equation}
\begin{cases}
I_{1}(t)= & I_{1}(0)\, e^{-\frac{c}{L}\mu_{1}t}\\
I_{2}(t)= & I_{2}(0)\, e^{-\frac{c}{L}\mu_{2}t}
\end{cases}\ .\label{eq:3.5}
\end{equation}

To measure light decay times, the experimental procedure consists in recording the RF discharge with a fast detector, (photomultiplier Hamamatsu H7827012) loaded on a 1 $k\Omega$ impedance, after a rapid switch off of the RF discharge. The switching-off operation must be much faster than the laser decay time. In our setup we obtained a sufficiently rapid switch off by grounding one of the two electrodes of the radio-frequency discharge by means of a mechanical switch. A validation of this technique is obtained by measuring the decay time of the plasma fluorescence which results to be of the order of few microseconds. Finally, we performed an exponential fit of the collected data. 

\subsection{Calibration of Intensities in Lamb Units}

To get accurate estimates of the Sagnac frequency, the light intensities input of the Extended Kalman filter must be calibrated in Lamb units \cite{Aronowitz}. To this aim, we propose an experimental method based on the observation of the birth of additional longitudinal modes of the laser while rising up the laser excitation power. This dynamical change is known as ``multimode transition'', and has been widely studied in the literature, mainly for medium size and large size ring laser \cite{MM,MM1}. The value of the mean light intensity for the multimode transition expressed in Lamb units $I_{th},$ is commonly defined multimode threshold. Different calculations of the multimode threshold were proposed \cite{MM1} which take into account only the plasma dispersion function, evaluated at frequencies of fundamental and higher order modes. To increase the accuracy of the multimode threshold it is convenient to account for cavity losses and back-scattering in the balance of gained and lost photons for longitudinal lasing modes. 

The threshold condition for multimode transition can be calculated from the stability analysis of the RL system. Starting from Eqs.(\ref{eq:2.1}), we can write the following system of equations for the intensities $I_{1,2}$ of two fundamental modes 

\begin{equation}
\mathbf{\dot{I}}=\text{Re}\left(\mathcal{\mathscr{D}}(\mathbf{E}^{*}) A \mathbf{E}-\mathcal{\mathscr{D}}(\mathbf{I}) B \mathcal{\mathscr{D}}(\mathbf{E}^{*})\mathbf{E}\right)\ ,\label{eq:3.6}
\end{equation}

where $\mathbf{I}=\left(I_{1},\, I_{2}\right)^{T}.$ If a new $m-$mode (lasing at $\omega_{1,2}+mc/L$) adds to the system, its dynamics will depend on $4$ intensities of light $I_{1,2,3,4}.$ The evolution of the system is still ruled by Eqs(\ref{eq:3.6}), provided that we substitute the $2\times2$ matrices $A$ and $B$ with the corresponding $4\times4$ matrices, calculated by means of the plasma dispersion function for 4 intensities (see Appendix).

We now consider the system in the initial configuration $\left(I_{1},I_{2},0,0\right)^{T}$
and look for the condition of the $m$-th mode growth. The lower diagonal block of the Jacobian matrix of the 4 modes dynamical system is given by 

\begin{equation}
\left(\begin{array}{cccc}
\cdots\quad\  & \cdots\quad\  & \cdots & \cdots\\
\cdots\quad\  & \cdots\quad\  & \cdots & \cdots\\
0\quad\  & 0\quad\  & \alpha_{3}-\theta_{31}I_{1}-\theta_{32}I_{2} & 0\\
0\quad\  & 0\quad\  & 0 & \alpha_{4}-\theta_{41}I_{1}-\theta_{42}I_{2}
\end{array}\right)\ ,\label{eq:3.7}
\end{equation}

where $\alpha_{3,4}$ are the gain minus losses of the clock-wise and counter-clock-wise $m$-th
mode, and $\theta_{31,41,32,42}$ are the cross saturation coefficients between clock-wise and counter-clock-wise modes of fundamental and $m$-th mode. By the Lyapunov linearization theorem \cite{key-6}, if the eigenvalues of the above matrix lie in the strictly positive complex half-plane, the equilibrium point is unstable, and the new laser modes can start to grow. Therefore a higher $m-$mode can be excited if $\alpha_{3}-\theta_{31}I_{1}-\theta_{32}I_{2}>0$ and $\alpha_{4}-\theta_{41}I_{1}-\theta_{42}I_{2}>0.$ Since the frequency difference of the two counter-propagating beams is much smaller than the Doppler width (see Appendix \ref{sec:Appendix}), we have $\theta_{31}\sim\theta_{42}\sim\theta_{ms},$
$\theta_{32}\sim\theta_{41}\sim\theta_{mc}\,$. Thus, to derive a threshold condition for the multimode operation we can take the average of the two inequalities and write

\[
\bar{\alpha}_{m}>\bar{\theta}_{m}\overline{I}\;,
\]

where $\bar{\alpha}_{m}=(\alpha_{3}+\alpha_{4})/2$, $\overline{I}=(I_{1}+I_{2})/2$ and $\bar{\theta}_{m} = (\theta_{ms}+\theta_{mc})/2$, and so the threshold condition for the intensity $I_{th}$ is

\begin{equation}
\bar{\alpha}_{m}=\bar{\theta}_{m}I_{th}\quad.\label{eq:3.8}
\end{equation}

This condition is not sufficient to determine the multimode threshold because active laser parameters depend on the value of the single pass gain at the threshold $G_{th}$, which is also unknown. However,
we can add to Eq. (\ref{eq:3.8}) a second equation representing the balance of the mean intensity for the fundamental modes, taking into account the average $(\hat{\mu}_1+\hat{\mu}_2) /2$ in Eq.(\ref{eq:2.9}). Therefore the system of equations in the variables $G_{th}$ and $I_{th}$ reads

\begin{equation}
\begin{cases}
G_{th}\, z_{m}^{(0)}(\xi)-\bar{\mu} & =2I_{th}G_{th}\left[z_{m}^{(2)}(\xi_{m})\right]  \\ 
G_{th}\, z^{(0)}(\xi)-\bar{\mu} & =I_{th}G_{th}\left[z_{s}^{(2)}(\xi)+z_{c}^{(2)}(\xi)\displaystyle{\frac{1+3\delta_{I}^{2}}{1-\delta_{I}^{2}}} \right] 
\end{cases},\label{eq:3.9}
\end{equation}

where $\xi_{m}$ is the normalized detuning averaged over the beams $3$ or $4,$ $\delta_{I}=(I_{1}-I_{2})/(I_{1}+I_{2}),$ $\overline{\mu}=(\mu_{1}+\mu_{2})/2,$ $z_{m}^{(0)}$, $z_{s}^{(0)}$ and $z_{c}^{(0)}$ are polarization contributions from the plasma dispersion function which are computed
in the Appendix. As m-modes are very close in frequency, their losses can be assumed with good approximation to be equal. The quantity $\delta_{I}$ can be estimated from the acquired intensity channels $V_{1,2}$ as $\delta_{I}=(V_1-V_2)/(V_1+V_2)$. It is worth noticing that the measure of $\delta_{I}$ is independent of multiplicative change of scale. The multimode condition can be derived by solving Eq.(\ref{eq:3.9}) 

\begin{equation}
\begin{cases}
I_{th} & =\frac{\Delta z^{(0)}\left(1-\delta_{I}^{2}\right)}{\Delta z^{(2)}+\delta_{I}^{2}\left(\Delta z^{(2)}-4z_{c}^{(2)}(\xi)\right)}\\
G_{th} & =\frac{\bar{\mu}}{z_{m}^{(0)}(\xi)-I_{th}\, z_{m}^{(2)}(\xi_{m})}
\end{cases},\label{eq:3.10}
\end{equation}

where $\Delta z^{(0)}=z^{(0)}(\xi)-z_{m}^{(0)}(\xi),$ $\Delta z^{(2)}=z_{s}^{(2)}(\xi)+z_{c}^{(2)}(\xi)-2\, z_{m}^{(2)}(\xi_{m})$ represent the difference between the fundamental and $m$-modes of
the $0^{th}$ and $2^{nd}$ population inversion contributions. The resulting value $I_{th}$ provides the desired calibration for $\{I_{1,2}(n)\}$ from the ADC acquired voltages$\{V_{1,2}(n)\}$ to Lamb units.
In addition, we have also derived an estimate of $G_{th}$ that will be used as the initial value of the gain monitor.

\subsection{Gain Monitor}

The intensity of the plasma fluorescence line at $632.8$ nm provides a good observable for monitoring the relative variations of the atomic population in the upper laser level. The calibration of the monitor signal is obtained by performing intensity steps in the neighborhood of the monomode working regime of the ring laser, exploiting the identification procedure described in Eq.(\ref{eq:2.9}), and a linear least squares fit. In fact, the second equation of the system in Eq.(\ref{eq:3.9}), representing the balance among gain, losses and mean intensities, holds for any value of G and $\overline{I}$. By solving this equation for the variable $G$, and using N measurements $\{\overline{I}(n)\}$ and $\{\delta_{I}(n)\}$ $(n=1,2,3\ldots,N)$ in the monomode regime, we obtain N estimates of the gain signal  

\begin{equation}
G(n)=\frac{\bar{\mu}}{z^{(0)}(\xi)-\overline{I}(n)\left(z_{s}^{(2)}(\xi)+z_{c}^{(2)}(\xi)\frac{1+3\delta_{I}^{2}(n)}{1-\delta_{I}^{2}(n)}\right)},
\end{equation}

where the mean losses value $\bar{\mu}$ is supposed to be constant and equal to the mean of RDT estimation. 

To account for the experimental setup, we can consider a simple linear measure model $G=a\, V_{p}+b$
for the gain monitor signal $V_{p}$, where the constants $a$ and $b$ have to be estimated by the linear least squares fit 

\begin{equation}
\left(\begin{array}{c}
\widehat{a}\\
\widehat{b}
\end{array}\right)=\text{argmin}_{a,b}\left\Vert \left(\begin{array}{cc}
V_{p}(1) & 1\\
\vdots & \vdots\\
V_{p}(N) & 1
\end{array}\right)\left(\begin{array}{c}
a\\
b
\end{array}\right)-\mathbf{G}\right\Vert ^{2},
\end{equation}

where $\mathbf{G}=[G(1),\ldots,G(n)]$ is the vector of gain estimates. The estimated constants $\widehat{a}$, $\widehat{b}$ and $G_{th}$ allow us to monitor the laser single pass gain by acquiring the signal $V_p$ without affecting the continuous operation of a RL. 

\section{Experimental Apparatus \label{sec:Experimental-Apparatus}}
\subsection{Mechanical mounting}
\begin{figure}
\includegraphics[width=8cm]{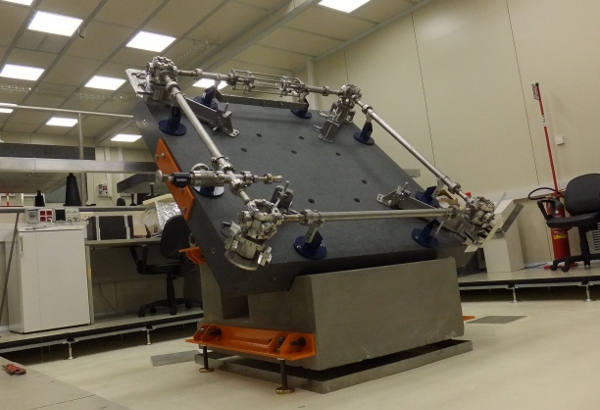}
\caption{\label{monum} The ring laser G-PISA attached to its monument, and pointing in the direction of the North Pole.}
\end{figure}

The ring laser G-PISA consists of a square optical cavity with a sidelength of 1.350 m. The four cavity mirrors are  contained in a steel vacuum chamber entirely filled with a He-Ne gas mixture. Two optically transparent windows are mounted on each corner of the cavity and allow to measure the 8 beams emitted by the cavity. In this work the G-PISA cavity has been fixed to a granite table that, in its turn, is firmly attached to a special monument completed in 2012. The monument is made of reinforced concrete, is oriented toward the local North direction and allows to hold the granite table with a tilt angle equal to the latitude of the laboratory. The concrete monument has two niches located on two opposite sides (one visible in the figure) for the housing of seismic equipment (a broad-band seismometer and a tiltmeter). The positioning of  monument has been done by using topographic references (angle with the north) and an inclinometer (angle with the local vertical). The estimated error in the monument
orientation was estimated to be less than $<1$$\: \rm{deg}$. In this configuration, the Sagnac frequency due to the Earth rotation is maximum and the rotation rate of the instrument is very stable in time. A picture of the G-PISA assembly, already oriented at the maximum signal, is shown in Fig. \ref{monum}. 
This is the best condition for reducing the environmental disturbances,  and for the systematic
study of the laser dynamics and control. 

\begin{figure}
\includegraphics[width=7cm]{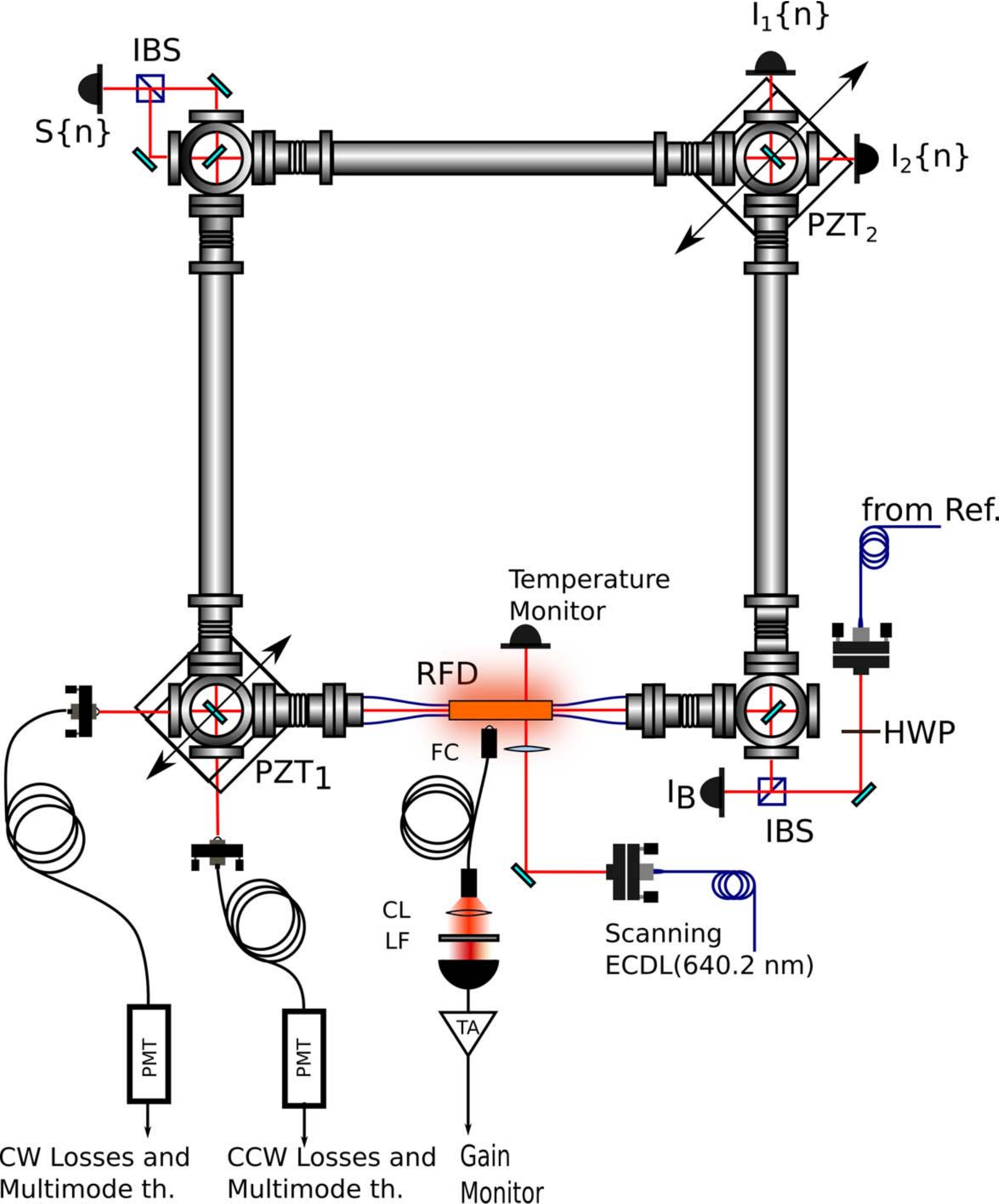}
\caption{\label{apparato}The experimental setup for the rotation measurement 
and the calibration of the ring laser parameters. PMT: photomultiplier tube. 
TA: transimpedance Amplifier, LF:Line Filter. CL: Collimating Lens. TA Gain for G-PISA is 1 $\rm{G \Omega}$. ECDL: Extended Cavity Diode Laser.}
\end{figure}

\subsection{Measurement setup} The optical setup is shown in Fig. \ref{apparato}. In the middle of one side of the cavity a pyrex capillary is mounted, 4 mm in internal diameter and 150 mm in length. The capillary plays a double role: it acts as a diaphragm selecting the $TEM_{00}$ spatial mode and it
gives the possibility to apply a radiofrequency electric field for the excitation of the  He-Ne plasma.
The RL G-PISA is operated with a gas mixture of $50\%$ $^{20}\rm{Ne}$ and $50\%$ $^{22}\rm{Ne},$ at a total pressure of $7.5$ mbar. The measurement of the Sagnac interferogram is obtained by combining the two output beams exiting one corner trough an intensity beam splitter while the single beam intensities are directly detected at the output of the adjacent corner. The Sagnac signal and the single beams are detected with a large area (5.8 $\times 5.8\, \rm{mm^2}$) Si photodiodes (Hamamatsu  S1227-66BR) followed by a transimpedance amplifier (FEMTO LCA-4K-1G) with a gain 
of $10^9\, \Omega$ and bandwidth of 4 kHz. One the corner opposite to the one for the Sagnac detection an optical beat setup is mounted between the clockwise beam and the $I_2$ stabilized He-Ne reference laser. The beat is detected by an avalanche photodiode whose current is amplified by a transimpedance amplifier  (FEMTO HCA-400M-5K-C) with a gain of 4 $\rm{k\Omega}$ and a bandwidth of 400 MHz. During G-PISA operation, the detuning of the clock-wise wave is kept constant
by a perimeter stabilization loop \cite{G-PISA,G-PISA1}, acting on the position of two opposite mirrors of the cavity. The ring laser frequency is locked in this way to the value where $z^{(0)}(\xi_{1,2})$ attains its maximum.

\subsection{Diagnostic apparatus}
To increase the level of precision and accuracy in the rotation-rate measurement, cold cavity and active medium parameters (Neon atomic kinetic temperature, total homogeneous broadening, and  isotopic composition) should be directly measured on the experimental apparatus. In order to measure the ring-down time of both the clockwise and counterclockwise modes, the beams exiting the corner opposite to the one dedicated to the intensities monitor are detected by two fiber-coupled photomultipliers (Hamamatsu H7827012). Two diagnostics have been arranged for the estimation of the active medium parameters: a fluorescence monitor for the gain variations and a laser probe interrogating the plasma through the pyrex capillary. 

\subsubsection{Population inversion monitor}
In order to perform an on-line measurement of the laser gain, we coupled part of the plasma fluorescence to a multi-fiber bundle. The collected light containing all the spectral contribution of the He-Ne discharge, is filtered by a line filter $1$ nm wide around $632.8$ nm and detected with a photodiode.
The photocurrent is amplified with a transimpedance stage with a gain of $1$ $\rm{G \Omega}$. The voltage $V_{p}$ of the photodiode is used as an optical monitor of the laser gain by recording the dependence of the output powers $I_{1}$ and $I_{2}$ on $V_{p}$, after losses have been estimated. 

\subsubsection{Spectroscopic probe of the gain medium}

Essential information about the gain medium can be extracted by observing the Doppler absorption of the plasma at 640.2 nm (the strongest closed optical transition of Ne). We setup a frequency tunable ECDL (Extended Cavity Diode Laser) crossing the He-Ne plasma through the pyrex capillary \cite{pdffitbib}. From this measurement one can get a precise estimation of the Doppler broadening, as well as of the isotopic composition of the gas. An example of this measurement is given in Fig. \ref{fig:A2}, where a ``standard'' He-Ne gas mixture has been used. The plasma temperature of G-PISA has been experimentally estimated as $T_{a}=(360\,\pm12)$ K. 

\section{Results and Discussion\label{sec:Results-and-Discussion}}

We  implement estimation and calibrations routines for G-PISA with models and techniques described in Sec.~\ref{sec:Ring-Laser-Dynamics} and \ref{sec:Calibration-Procedures}, respectively. In addition, we also implemented a data quality criterion that discards large outliers due to electronic spikes. 

\begin{figure}
\includegraphics[width=9cm]{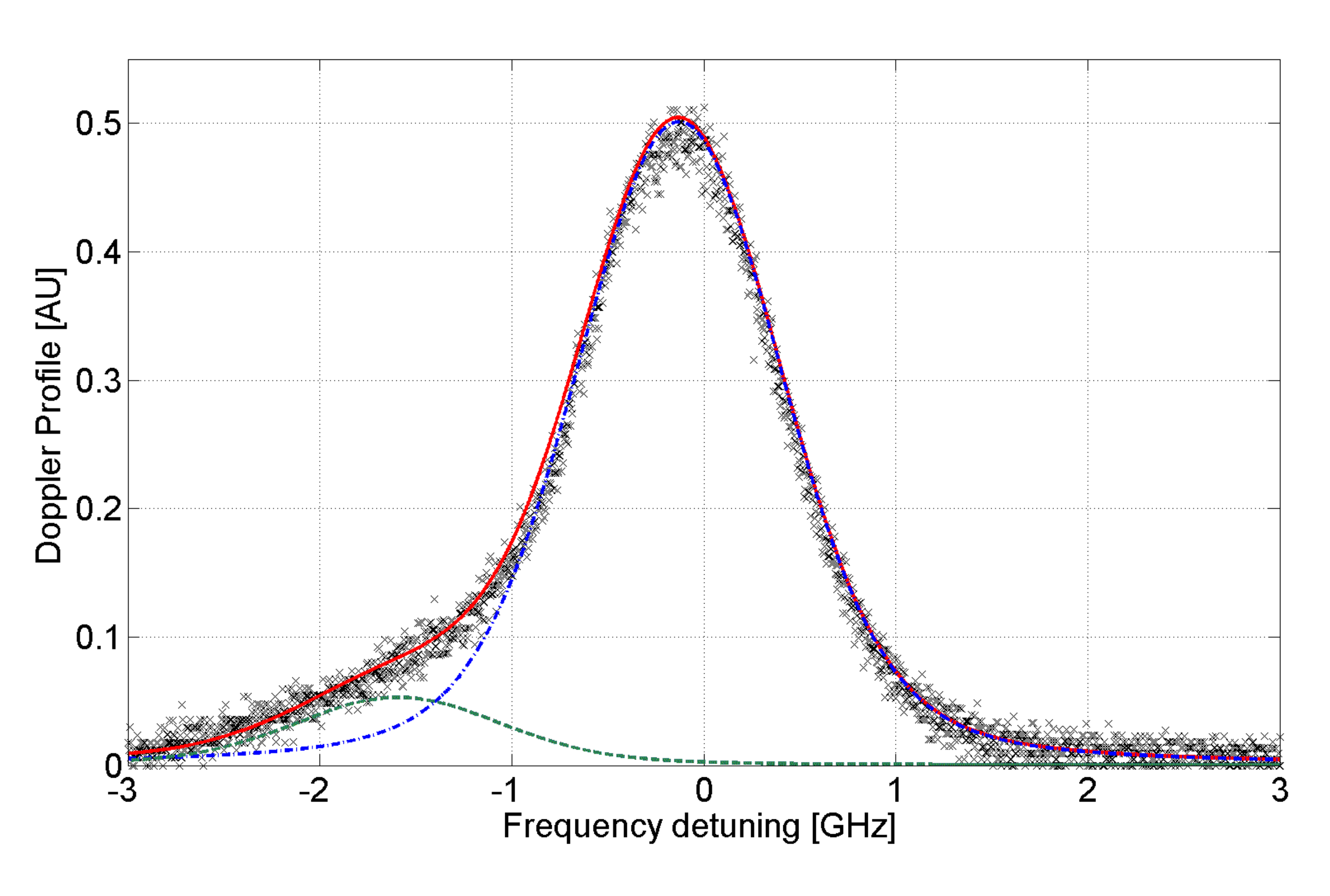}
\caption{\label{fig:A2} Absorption profile of the closed optical transition in Neon at 640.2 nm, allowing for the Neon temperature estimation. The measurement is taken in typical operation conditions for a plasma of $He-Ne$ standard mixture at $4.5$ mbar. Using the standard Matlab fitting procedure and the function $z^{(0)}(\xi)$, we get a reduced R-squared of $0.9947$ and the fitting parameters $a'=-4.9 \pm 0.1$, $b'=-0.2 \pm 0.05$, $\eta'=0.27 \pm 0.02$, $b''=-2.44 \pm 0.05$, and $k''/k'= 0.11 \pm 0.03$.}
\end{figure}

To get an estimation of the plasma temperature, gas pressure and isotopic concentration, we fit the normalized measures of a laser diode on the standard $He-Ne$ gas mixture  in Fig.\ref{fig:A2} to the function $z^{(0)}(\xi)$ in Eq.(\ref{eq:A5}). To account for detuning uncertainties in the laser probe, we scaled the experimental abscissa so that $\xi' = a' \xi + b'$, $\xi'' = a'' \xi + b''$. The fit parameters $a'$, $b'$, $\eta'$, $k'$, and $b''$, $k''$, are related to $^{20}$Ne, and $^{22}$Ne isotopes, respectively; the remaining parameters of $^{22}$Ne isotope are calculated as  $a''=\sqrt{22/20}a'$, and $\eta'' = \sqrt{22/20} \eta'$. From the fit results, we get $ T_{Ne} = (360 \pm 12)$K.

\begin{figure}[h]
\includegraphics[width=9cm]{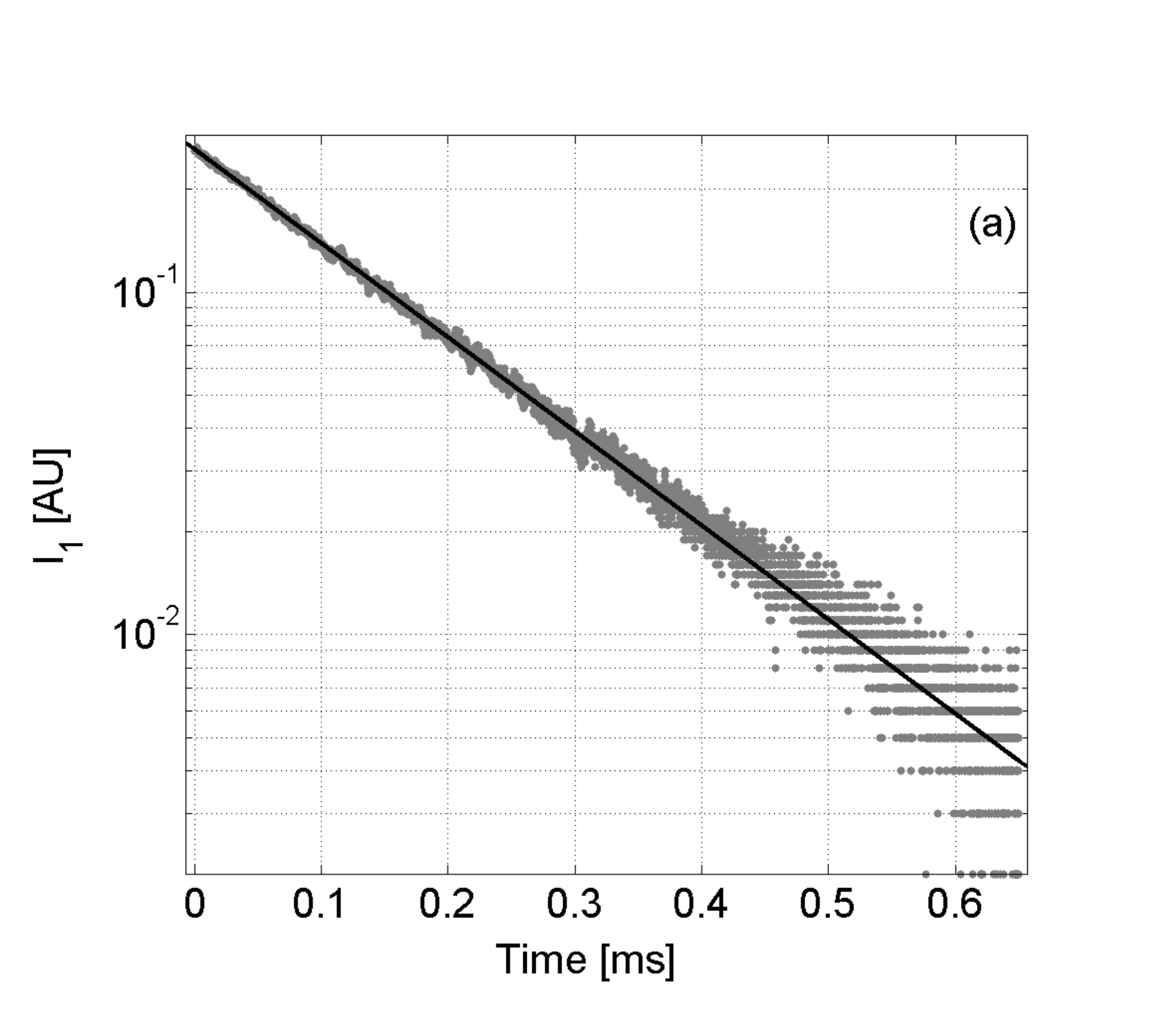}
\includegraphics[width=9cm]{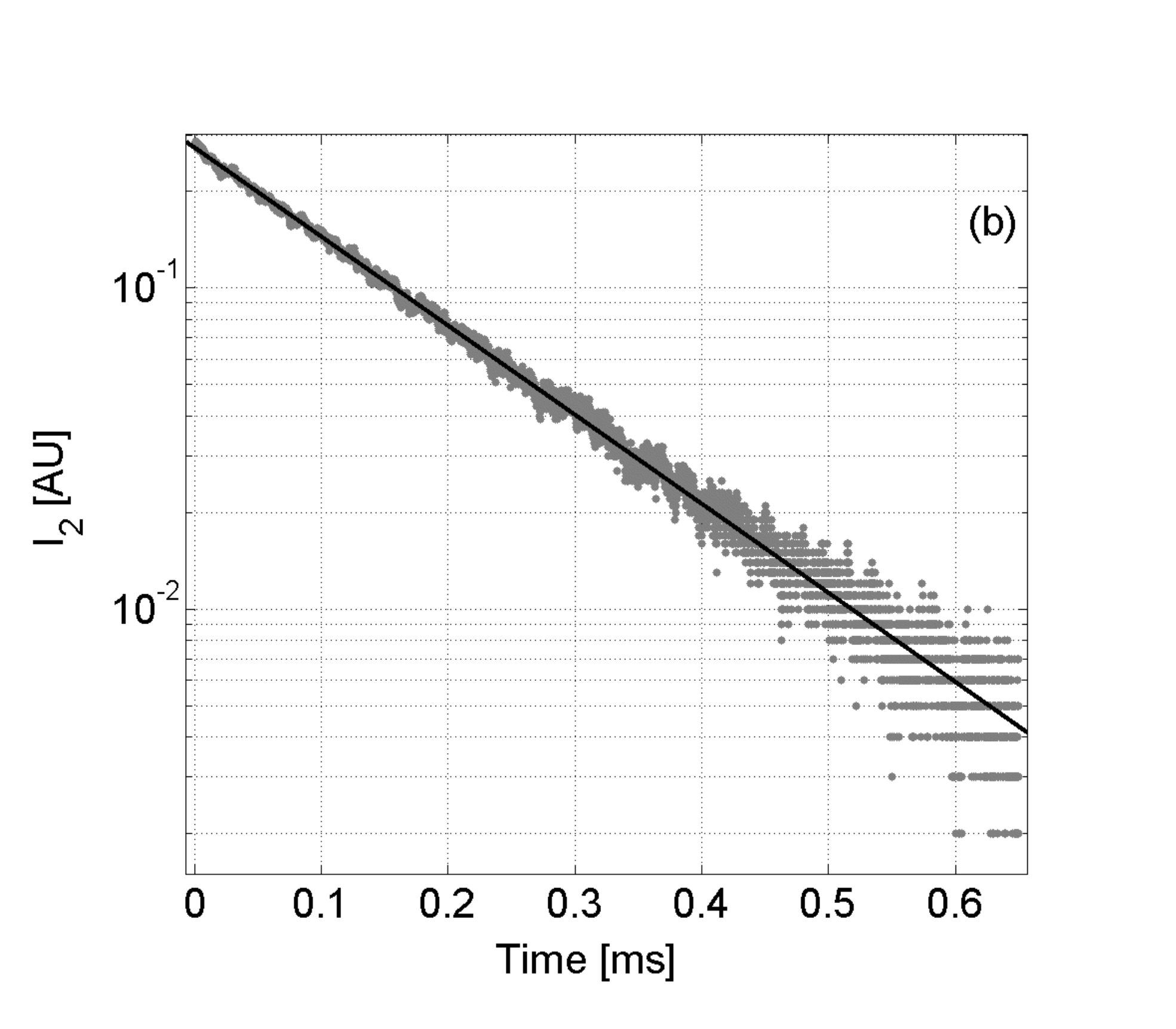}
\caption{\label{fig:1} Plot of RDT data sampled at $10\,\text{MHz}$ with the fitting functions $n_{1,2}\exp(-\mu_{1,2}ct/L)$ for beams $1$ $(a)$, and $2$ $(b)$, respectively. The reduced $R-$squared from Matlab fitting toolbox are $R_{1}^{2}=0.9969,\,$and $R_{2}^{2}=0.9955,\,$ and the fitting parameters are $\mu_{1}=\left(1.136\pm0.02\right)\times10^{-4},$ $\mu_{2} = \left(1.146\pm0.02\right)\times10^{-4},$ $n_{1}=\left(1.898\pm0.005\right)\times10^{-1},$
$n_{2}=\left(1.982\pm0.005\right)\times10^{-1}.$ }
\end{figure}

In Fig. \ref{fig:1} we report the results of the ring down time measurements, the losses for the two beams were estimated as $\mu_{1}=\left(1.136\pm0.02\right)\times10^{-4}$, $\mu_{2} = \left(1.146\pm0.02\right)\times10^{-4}$. Note that within the precision of the fit it is $\mu_1 \sim \mu_2.$
 
\begin{figure}
\includegraphics[width=9cm]{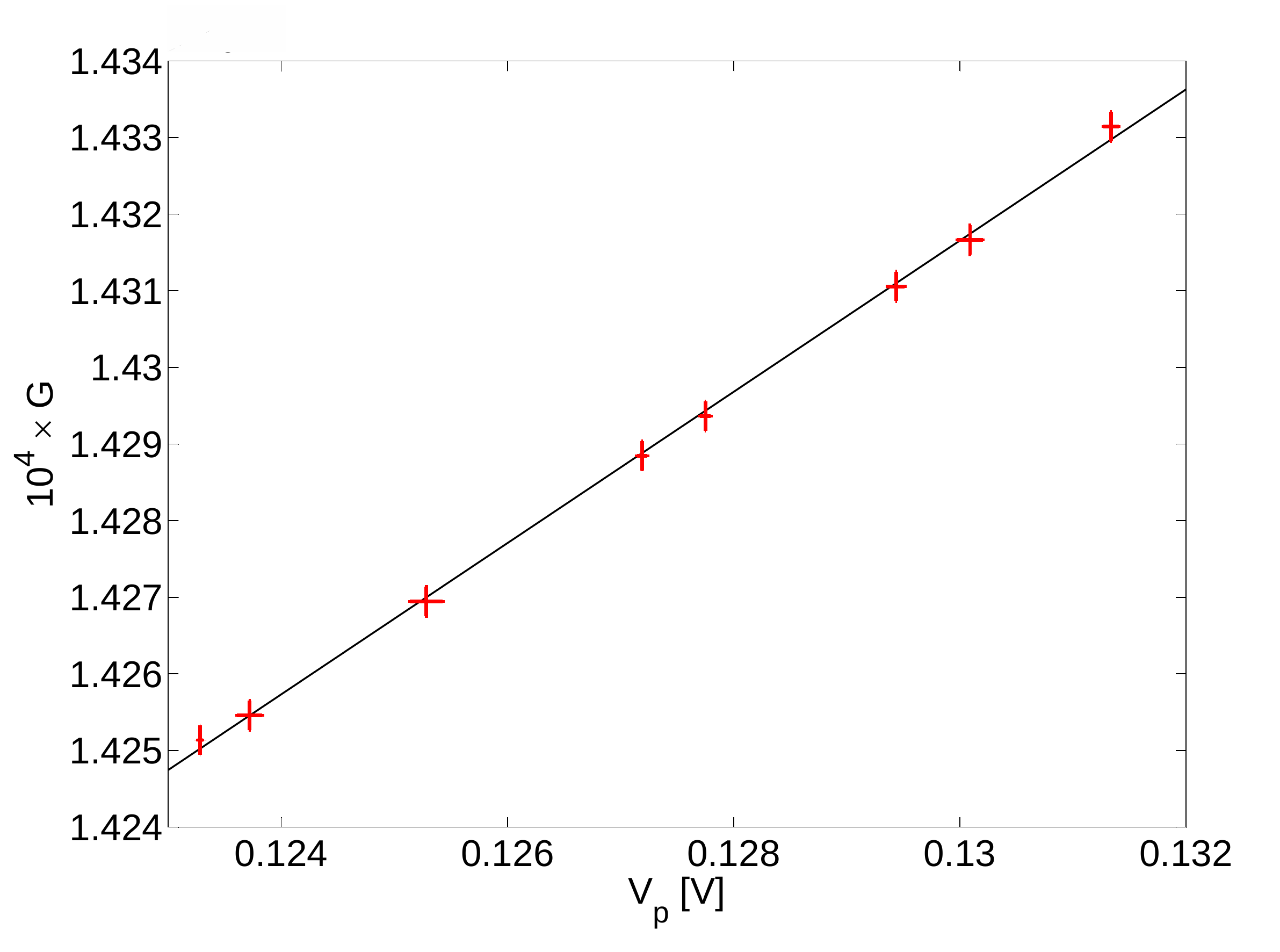}
\caption{\label{fig:gainm} Plot of the single pass gain $G$ of G-PISA as a function of the gain monitor $V_{p}$. Each point represents the average of 10 measurements and the corresponding error bar is their standard deviation. The linear fit gives $\widehat{a}=9.87 \times 10^{-5}$, $\widehat{b} =1.303\times10^{-4}$ $\sigma_a=1.57\times 10^{-6}$, and $\sigma_b=2\times10^{-7}$.}
\end{figure}

\begin{table}
\centering
\begin{tabular}{|c|c|}
\hline 
Error Source & Freq. error \\
\hline
Back-scattering $ \mathcal{R}_{2} e^{-X}+\mathcal{R}_{1} e^{X}  + $ c.c.  & $0.4695$ Hz \\
\hline  
Null Shift $\tau\,(I_{1}-I_{2})$ & $-8.7 \times 10^{-4}$ Hz \\
\hline 
Atomic Scale Factor $\sigma_{1}-\sigma_{2}$ & $5.56\times10^{-6}$ Hz \\
\hline
Cross Dispersion $I\,(\tau_{21}-\tau_{12})$ & $1.75\times10^{-6}$ Hz \\
\hline 

\end{tabular}
\caption{Contributions to the accuracy budget of G-PISA from systematic errors in the estimate
of Lamb parameters.}
\end{table}

Once the calibration of the experimental apparatus has been  performed, we used the monobeam intensity offsets, modulation amplitudes and phases,  and the gain monitor to estimate both cold cavity and active medium parameters. The intensities $I_{1,2}(t)$ and interferogram $S(t)=|E_1+E_2|^2 $,  sampled at $5$ kHz, were collected in two days. We estimated the Sagnac frequency by means of the EKF \cite{Noi}, with deterministic dynamics given by  Eqs.(\ref{eq:2.6}), and with measure vector

\begin{equation}
\mathbf{y} = \left( \begin{array}{c} 
I_1(t)\\
I_2(t)\\
S(t)\\
\end{array} \right) \ , 
\end{equation}

where $S(t)\simeq$ $ \sqrt{I_1(t)I_2(t)} \cos(Im(X(t)))$ neglecting the contribution $I_1(t) + I_2(t) $. Finally, the performances of the EKF routine were compared with AR2, the standard frequency estimation algorithm for RL beat note. 

\begin{figure}
\includegraphics[width=9cm]{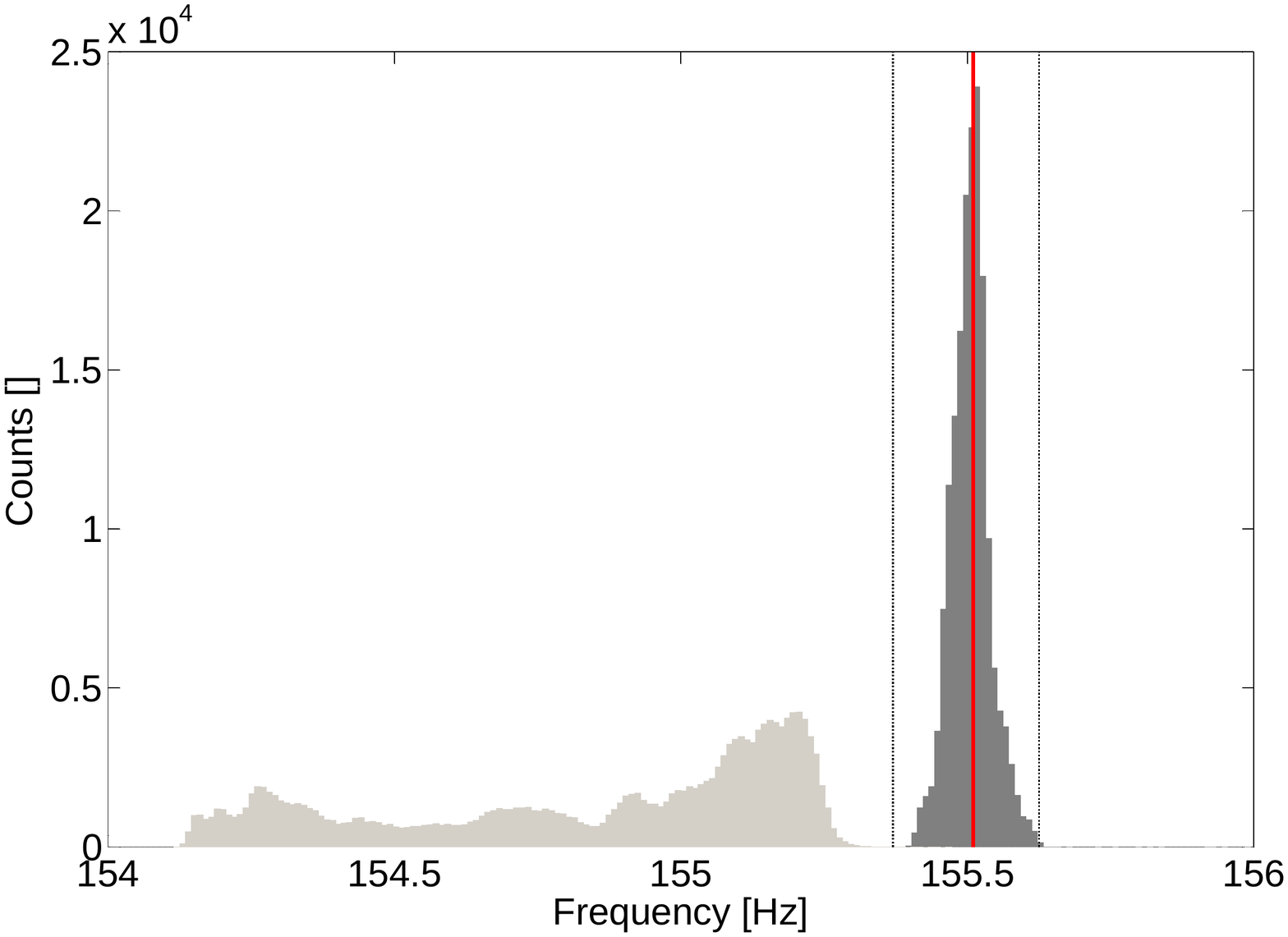}
\caption{\label{fig:hists}Histograms of the estimates of AR2 (pale gray) and EKF (dark gray) during 2 days of G-PISA data. The red line is the expected Sagnac frequency due to Earth rotation, and the dotted lines represent its residual uncertainty bounds due to geometric and orientation tolerances.}
\end{figure}

Fig. \ref{fig:hists} shows the histograms of the AR2 and EKF estimates.  Note that the EKF mean is shifted with respect to AR2 mean (effective removal of the  frequency null shift), and that the EKF standard deviation is $\sim 10$ times smaller than the standard deviation of AR2 estimator. The G-PISA long-term stability and accuracy have been so increased.

\begin{figure}
\includegraphics[width=9cm]{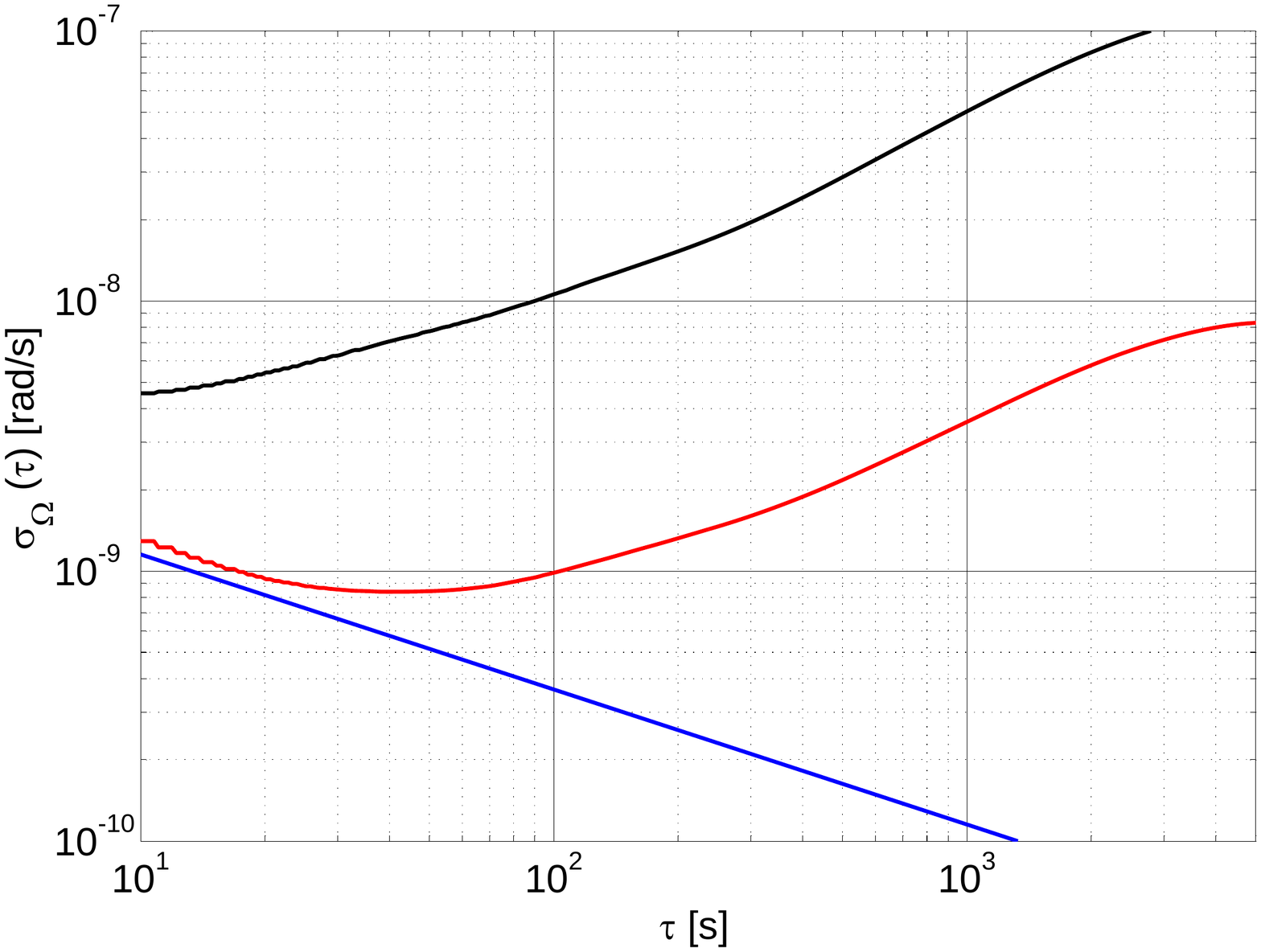}
\caption{\label{fig:avar_final} Allan variances of AR2 (upper curve) and EKF
(lower curve) rotational frequency estimates. The straight line represents the shot noise level of G-PISA for a cavity quality factor of $5.4\times10^{11}$ and output power of $4$ nW.}
\end{figure}

Moreover, in Figure \ref{fig:avar_final} we plot the Allan standard deviation of the AR2 and EKF estimates, and the expected Allan deviation curve of a frequency signal corrupted by shot noise. 
 
\section{Conclusions\label{sec:Conclusions}}

We have thoroughly studied the RL dynamics in order to implement identification and calibration methods for the cold cavity and active medium parameters. The identification method is based on the first harmonic approximation of the steady state solution of RL equations and the minimization of a quadratic functional over the Hilbert space of periodic signals. On the other hand, the dynamics of a laser with cavity detuning shows many monomode or multimode dynamical behaviors  that can be exploited to get rid of systematic errors of Sagnac frequency estimation.  The calibration method presented in this paper is based on the measure of the threshold of the multimode transition and the plasma dispersion function. The systematic errors associated with the latter measurements dominate over statistics errors, as plasma parameters depend also on RF discharge details, e.g. shape of the capillary discharge. However, the accuracy of the identified cold cavity parameters  depends on the amount of losses and backscattering of light: higher quality mirrors lead to potentially higher accuracy of the estimated Sagnac frequency. Much work has still to be devoted to improve the calibration procedure for the application of RL to fundamental physics. The problem of pushing calibration and identification methods to their intrinsic accuracy limit, with or without the addition of a calibrated rotation signal to the ring laser input, would deserve further investigation and it will be the topic of a forthcoming paper. 

\section*{Acknowledgments}

The authors gratefully acknowledge the support of  Giorgio Carelli and Enrico Maccioni during the setup and operation of the G-PISA apparatus.  Its also a pleasure to acknowledge Filippo Bosi for the project and realization of the concrete monument, and for the alignment of G-PISA to the north pole direction.

\clearpage

\appendix

\section*{Appendix: Calculation of cross-saturation \label{sec:Appendix}}

We adress to the problem of calculating the cross-saturation coefficients $z_{ms}^{(2)}(\xi),\, z_{mc}^{(2)}(\xi)$ for two multimode counter-propagating waves $E_{3,4}$, lasing at $\xi_{3,4}=\xi_{1,2}+nc/L,$ $n\in\mathbb{Z}$, that arise at multimode transition of a RL. 

The polarization components for the waves $E_{3,4}$ at multimode threshold is given by Eq.(\ref{eq:A1}), provided that the subscript $1,2$ is substituted with $3,4\,,$ and $\rho^{(2)}(v,E_{1,2})$ is replaced with the following expression 
\begin{widetext}

\begin{equation}
\rho_{m}^{(2)}(v,E_{1,2})=\frac{N\, e^{-\frac{v^{2}}{\Gamma^{2}}}}{2\gamma_{a}\gamma_{b}\hbar\Gamma}\left(1-\frac{\gamma_{a}+\gamma_{b}}{\gamma_{ab}}I_{1}\frac{1}{1+(\xi_{3}-f_{m}+v/u)^{2}}-\frac{\gamma_{a}+\gamma_{b}}{\gamma_{ab}}I_{2}\frac{1}{1+(\xi_{4}-f_{m}-v/u)^{2}}\right)\quad\label{eq:A7}
\end{equation}

where $f_{m}=nc/(L\Gamma)$. The expression for the multimode atomic polarization reads 

\begin{equation}
\mathcal{P}^{(3)}(E_{3,4})=\frac{\sqrt{\pi}A\mathcal{Z}_{I}\left(0\right)}{\gamma_{ab}\gamma_{a}\gamma_{b}}E_{3,4}\left(z^{(0)}(\xi_{3,4})-z_{ms}^{(2)}(\xi_{3,4})I_{1,2}-z_{mc}^{(2)}(\xi)I_{2,1}\right)\quad,\label{eq:A8}
\end{equation}

where 

\begin{equation}
\begin{cases}
z_{ms}^{(2)}(\xi_{3,4})= & 2\eta^{2}\frac{\gamma_{a}+\gamma_{b}}{\gamma_{ab}}\frac{\mathcal{Z}\left(\xi_{3,4}-f_{m}\right)-\mathcal{Z}\left(\xi_{3,4}\right)^{*}}{f_{m}(if_{m}+\eta)\mathcal{Z}_{I}\left(0\right)}\\
z_{mc}^{(2)}(\xi_{3,4})= & \frac{\eta}{\xi-\frac{f_{m}}{2}}\frac{\gamma_{a}+\gamma_{b}}{\gamma_{ab}}\frac{(\xi-\frac{f_{m}}{2})\mathcal{Z_{I}}\left(\xi_{4,3}-f_{m}\right)-\frac{\eta}{2}\left(\mathcal{Z}\left(\xi_{4,3}-f_{m}\right)+\mathcal{Z}\left(\xi_{3,4}\right)^{*}\right)}{\mathcal{Z}_{I}\left(0\right)(\eta+i(\xi-\frac{f_{m}}{2}))}
\end{cases}\ .\label{eq:A9}
\end{equation}
\end{widetext}

In addition, for the cross saturation coefficients between $1$ and $2$ modes of fundamental and $m$-th mode, we have 

\begin{equation}
\begin{cases}
\theta_{31,42} & =\text{Re}\left[z_{s}^{(2)}(\xi_{3,4})\right]\\
\theta_{32,41} & =\text{Re}\left[z_{c}^{(2)}(\xi_{3,4})\right]
\end{cases}\quad.\label{eq:A10}
\end{equation}

When more than one isotope is present in the gas mixture, one must modifies Eqs.(\ref{eq:A9}) using the weighted average of each isotope contribution, as in Eqs.(\ref{eq:A6}). As a final remark, we note that, for large RL sensing the Earth rotation, the difference between the normalized detunings of  each m-mode is very small, e.g. $|z^{(0)}(\xi_{3})-z^{(0)}(\xi_{4})|\lesssim10^{-6}$ for G-PISA. Consequently, the following approximation holds

\[
\begin{cases}
z_{ms}^{(2)}(\xi_{3,4}) & \sim z_{ms}^{(2)}\left(\frac{\xi_{3}+\xi_{4}}{2}\right)\\ 
z_{mc}^{(2)}(\xi_{3,4}) & \sim z_{mc}^{(2)}\left(\frac{\xi_{3}+\xi_{4}}{2}\right)
\end{cases}\quad \].

\end{document}